\documentclass[12pt]{article}
\usepackage{graphicx}

\newsavebox{\sboxpubnumber}
\newsavebox{\sboxpubdate}
\newcommand{\pubdate}[1]{\begin{lrbox}{\sboxpubdate}{#1}\end{lrbox}}

\newcommand{\Title}[1]{\begin{center} {\Large #1 } \end{center}}
\newcommand{\Author}[1]{\begin{center}{ \sc #1} \end{center}}
\newcommand{\Address}[1]{\begin{center}{ \it #1} \end{center}}
\newcommand{\andauth}{\begin{center}{and} \end{center}}
\newcommand{\auth}{\begin{center}{} \end{center}}

\newenvironment{Abstract}{\begin{quotation}  }{\end{quotation}}
\newenvironment{Presented}{\begin{quotation} \begin{center}
             PRESENTED AT\end{center}\bigskip
      \begin{center}\begin{large}}{\end{large}\end{center}
      \end{quotation}}
\newcommand{\Acknowledgements}{\bigskip  \bigskip \begin{center} \begin{large}
             \bf ACKNOWLEDGEMENTS \end{large}\end{center}}

\newcommand{\dbar}{\not{\!{\!D}}}

\begin{document}

\begin{titlepage}
\pubdate{30/11/2001}                    

\vfill
\Title{Is inflatino production duirng preheating a threat to nucleosynthesis?}
\vfill
\Author{Rouzbeh Allahverdi\footnote{presented the talk, based on the
work in Ref. \cite{abm}.}}
\Address{Physik Department, TU M\"nchen \\
James Franck Strasse,
D-85748, Garching, Germany}
\vfill
\auth
\vfill
\Author{Mar Bastero-Gil}
\Address{Scoula Normale Superiore \\
Piazza dei Cavalieri 7, 56126 Pisa, Italy}
\andauth
\vfill
\Author{Anupam Mazumdar}
\Address{ICTP \\
Strada Costiera, 34014 Trieste, Italy}
\vfill
\begin{Abstract}
We discuss the production of inflatino the
superpartner of the inflaton due to vacuum fluctuations during
preheating and argue 
that they do decay alongwith
the inflaton to produce a thermal bath. Therefore they do not survive until
nucleosynthsis to pose a threat to it. 
\end{Abstract}
\vfill
\begin{Presented}
    COSMO-01 \\
    Rovaniemi, Finland, \\
    August 29 -- September 4, 2001
\end{Presented}
\vfill
\end{titlepage}
\def\thefootnote{\fnsymbol{footnote}}
\setcounter{footnote}{0}

\section{Introduction}
Inflation is perhaps one of the best paradigms of the early 
Universe which solves some of the problems of the standard 
Big Bang cosmology \cite{guth}. In addition, quantum fluctuations of the
inflaton field generated during inflation keep their imprint intact to
match the observed anisotropy in the present
Universe which is one part in $10^{5}$ \cite{bunn}.
Once inflation ends, the homogeneous inflaton field oscillates
coherently around the bottom of the potential. The inflaton
oscillations decay
when the Hubble parameter $H \sim
\Gamma_{\phi}$, where $\Gamma_{\phi}$ is the decay rate. The decay
rate essentially depends on the inflaton couplings to
other particles \cite{dolgov}.  Recently it has been realized that
there can be an explosive particle production due to non-perturbative
effects \cite{brand}. Effectively, the problem
turns out to be quantizing the bosonic and the fermionic fields in a
time-varying inflaton background. The production of bosons and
fermions differs in
its nature due to Pauli's exclusion principle, which prohibits
excessive production of fermions compared to their bosonic
counterparts \cite{ferm}.

The spin $3/2$ gravitino occurs in supersymmetric theories as a
superpartner of the graviton \cite{bailin}. A massless gravitino only
possesses $\pm 3/2$ helicity states. However, once supersymmetry is
broken the
gravitinos become massive, and they possess all four helicity states.
Soon after realizing that the helicity $\pm 3/2$ states of a
massive gravitino can be produced non-perturbatively \cite{anu}, it
was found out that the helicity $\pm 1/2$ states
can also be produced from vacuum fluctuations
\cite{kallosh,giudice,kallosh1}. They
are produced even more abundantly compared to helicity
$\pm 3/2$ states due to the Goldstino nature of helicity $\pm 1/2$
states which implies the absence of any Planck mass suppression in
their couplings.  

If supersymmetry is
required to solve the gauge hierarchy problem, then, in the gravity
mediated supersymmetry breaking, the gravitino gets a mass around,
${\cal O}(\rm TeV)$. The lifetime of the gravitino is
quite long, $\tau_{3/2} \sim {\rm M}_{\rm p}^2/m_{3/2}^3 \sim
10^{5}(m_{3/2}/TeV)^{-3}$sec, and hence its decay products can affect
light element abundances from big bang nucleosynthesis
\cite{sarkar}. This leads to a strong
constraint on the reheat temperature, $T_{\rm rh} \leq
10^{10}$ GeV, in order not to overproduce gravitinos in the thermal
bath \cite{ellis}. It also tightly constrains any non-perturbative
gravitino production during preheating.

Here we briefly discuss the decay of inflatinos and helicity $\pm 1/2$
gravitinos produced during preheating (details can be found in
Ref. \cite{abm}). We begin with an introduction of a supersymmetric
inflationary
model with a single multiplet, and then discuss decay rates of the
inflaton and its superpartner inflatino in two models: with
Planck mass
suppressed coupling, and with Yukawa couplings to the visible
sector.  We then establish an equivalence between the helicity $\pm
1/2$ gravitino coupling to the supercurrent and that of the
inflatino in the supergravity Lagrangian when the
amplitude of the inflaton oscillations are small compared to the
Planck mass. In the last section we give a qualitative discussion upon
the gravitino decay when more than one chiral fields are present.

\section{Inflaton decaying via gravitational coupling}

As a first example we consider a new inflation model proposed in
Ref.~\cite{ross} where the
inflaton sector and the visible sector interact only gravitationally. 
While setting the
cosmological constant at the global minimum to zero, the simplest form
of a superpotential emerges \cite{ross}

\begin{eqnarray}
\label{superpotential}
{\rm I} = {\Delta^2 \over {\rm M}} (\Phi - {\rm M})^2\,,
\end{eqnarray}
where $\Delta$ determines the scale of inflation, $\Phi$ is the
inflaton superfield, and $\rm M \equiv {\rm M}_{\rm P}/\sqrt{8 \pi}$ is the
reduced Planck mass. The amplitude
of the density perturbations produced during inflation is fixed by
COBE, leading to
$\Delta/{\rm M} \approx 5\times 10^{-3}$. The
scalar potential derived from the above superpotential has a form
\begin{eqnarray}
\label{potential}
V = e^{\sum_{j}\left(|\Phi_j|/{\rm M}\right)^2}\left(\sum_{k} \left |
\frac{\partial W_{\rm tot}}
{\partial\phi_k}+\frac{\phi^{\ast}_{k}W_{\rm tot}}{{\rm M}^2} \right
|^2 \, \right.  \nonumber \\ \left. -3\frac{|W_{\rm tot}|^2}{{\rm
M}^2}\right)\,,
\end{eqnarray}
where we have assumed minimal K\"ahler function and we consider the
total superpotential to be
\begin{eqnarray}
\label{totsuper}
W_{\rm tot}={\rm I}+{\rm L}\,,
\end{eqnarray}
where ${\rm L}$ represents the visible sector. The leading order term
in the scalar potential
generates trilinear coupling to the scalars in the visible sector with
a gravitational strength $\sim \Delta ^2/{\rm M}^2$, corresponding to
a decay rate $\Gamma_{\phi} \sim \left(\Delta ^6/{\rm
M}^5\right )$ for the inflaton \cite{ross}. Then the
reheat temperature of the Universe can be estimated by
\begin{eqnarray}
T_{\rm rh} \sim \left(\frac{30}{\pi^2
g_{\ast}}\right)^{1/4}\left(\Gamma_{\phi}{\rm M} \right)^{1/2} \approx
10^{-1} \frac{\Delta^3}{{\rm M}^2}\,.
\end{eqnarray}
For $\Delta/{\rm M} \sim 5 \times 10^{-3}$, the reheat temperature is
around $T_{\rm r} \sim 10^{10}$ GeV.

The equation of motion for the helicity $\pm 1/2$ gravitino in a
cosmological background has been derived in the literature by using
alternative approaches \cite{kallosh,giudice,kallosh1}.
The important realization is that when the amplitude of the
oscillations is much smaller than the reduced Planck mass, the
equation of motion for the helicity $\pm 1/2$ gravitino is effectively
that of the goldstino (which is the inflatino up to a phase) in a
global supersymmetric limit \cite{kallosh,giudice,kallosh1}.
The evolution of the
inflatino, which we define here as $\tilde{\phi}$, follows
\cite{giudice}
\begin{eqnarray}
\label{goldstino}       
i{\gamma}^{0}\dot{\tilde{\phi}} - \hat{k}\tilde{\phi} - m_{\rm
eff}\tilde{\phi} = 0\,,
\end{eqnarray}
where $\hat{k} = {\gamma}^{i}{k}_{i}$, and ${k}_{i}$ are components of
the physical momentum, while $\gamma^{i}$ are the gamma matrices.

When the amplitude of the inflaton oscillations $|\phi| \ll {\rm M}$,
the effective mass of the helicity $\pm 1/2$ gravitinos, for a single
chiral superfield and after a phase rotation, is simply the mass of
the fermionic component of the
inflaton field \cite{giudice}. This is the same as the mass of the
inflaton which is $\Delta^2/{\rm M}$ for the superpotential in
Eq.(3). On the other hand, the helicity $\pm 3/2$
gravitinos have a mass given by \cite{bailin}
\begin{eqnarray}
\label{m_3/2}
m_{\pm 3/2} \equiv e^{\phi^2/2{\rm M}^2}\frac{|{\rm I}|}{{\rm M}^2}
\sim \frac{\Delta^2} {{\rm M}}\left(\frac{\phi(t)}{{\rm M}}\right)^2
\,.
\end{eqnarray} 
This leads to a simple
inequality in various mass scales
\begin{eqnarray}
\label{sca}
m_{\phi}\approx m_{\pm 1/2} > H > m_{\pm 3/2}\,.
\end{eqnarray}

Now we analyze the decay rate of the inflatino. Consider the following
interaction found in the Lagrangian \cite{bailin}
\begin{eqnarray}
\label{dom}
|{\rm det}~e|^{-1}{\cal L}= -\frac{1}{2} e^{G/2} G^{i} G^{j} \bar
\chi_{i}{\chi}_{j{\rm L}} + {\rm h}.{\rm c}. \,,
\end{eqnarray}
where $G^{i}$ is the derivative of the K\"ahler potential with respect
to left and right chiral components. We can fix the index; $i= \phi$,
corresponding to the inflaton sector. This leaves the other index $j$
to run on the chiral components of the visible sector $\rm L$. It
turns out that the inflatino $\tilde \phi$ decays into scalars $\varphi$ and
fermions $\chi$ in the visible sector through terms like
$\Delta^2/{\rm M}^2 \tilde \phi \chi \phi$. This yields
\begin{eqnarray}
\Gamma_{\tilde \phi} \approx \frac{\Delta^6}{{\rm M}^5} \,.
\end{eqnarray}
This decay rate, not surprisingly, is the same as the decay rate of
the inflaton. Now, if we argue that the helicity $\pm 1/2$ states of the
gravitino essentially behave as inflatino when the
amplitude of the inflaton oscillations $|\phi | \ll {\rm M}$, then
, we may argue that the helicity $\pm 1/2$ gravitinos decay
alongwith the inflaton. Intuitively, this makes sense, because if
supersymmetry is restored at the bottom of the potential in the
absolute minimum, then only the $\pm 3/2$ components of the gravitino
should survive. However, to be
more concrete we must study the gravitino interactions.

The gravitino interaction terms appear from the couplings between the
gravitino field and the supercurrent
\begin{eqnarray}
\label{lag1}
{\cal L}_{\psi J}=\frac{1}{\sqrt{2}{\rm M}}\bar \Psi_{\mu}\dbar
 \varphi^{\ast j} \gamma^{\mu}\chi_{j {\rm L}} +\frac{i}{\sqrt{2}{\rm
 M}}e^{G/2}G^{i}\bar \Psi_{\mu}\gamma^{\mu}\chi_{i {\rm L}} \nonumber
 \\ + {\rm h.c.} \,,
\end{eqnarray}
where $\mu$ stands for the space-time index, $\chi_{i}$ is a fermionic
field and $\varphi^{i}$ is its bosonic superpartner. Here the
subscripts $i,j$ correspond to the visible sector ${\rm L}$, which
contains the light
degrees of freedom.  We have neglected the vector multiplets in the
above equation and assumed $\phi$ to be homogeneous. The total
derivative $D_{\mu}$is defined by
\begin{eqnarray}
D_{\mu} = \partial_{\mu} + \frac{1}{2} \omega_{\mu ab}\sigma^{ab}\,,
\end{eqnarray}
where $\omega_{\mu ab}$ is the spin connection.

The interaction terms
proportional to $\gamma^{\mu}\Psi_{\mu}$ are usually not necessary in
a static limit of the background field (i.e. inflaton field), because
$\gamma^{\mu}\Psi_{\mu}=0$ acts as a constraint for a gravitino field
in a static background. However, this need not be true in a non-static
background. It has been shown that in an expanding Universe, and in a
time-varying inflaton background, $\pm 1/2$ helicity states follow
$\gamma_{\mu}\Psi^{\mu}\neq 0$ \cite{kallosh}. Although, the same
constraint continues to hold good for the helicity $\pm 3/2$
components of the gravitino in the same background along with the
Dirac equation \cite{anu}. 

After several oscillations of the inflaton field $|\phi| \ll {\rm M}$,
or, equivalently $H \ll m$. All the fields whose effective mass is
larger than
the Hubble parameter during the oscillations of the inflaton would 
actually not feel any effect of curvature of the Universe. 
Therefore we replace $\pm 1/2$ helicity of the gravitino by an ansatz
\begin{eqnarray}
\label{ansatz}
\Psi_{\mu} \sim  \sqrt{\frac{2}{3}}\frac{\rm
M}{\rho_{\phi}^{1/2}}{\partial}_{\mu}\eta \,,
\end{eqnarray}
where $\eta$ represents the goldstino. At this moment this
prescription seems to be unwarranted, but, we shall see that this choice
of derivative wavefunction leads to the interactions of the helicity
$\pm 1/2$ gravitino to that of the inflatino. A similar expression has
beenpreviously used in Refs.~\cite{moroi}, where 
the authors have been studying the scattering processes of the 
helicity $\pm 1/2$ gravitino in a limiting case when the energy
scale of the gravitino is larger than its mass in a flat space-time.
Here, again we have a similar situation where the helicity $\pm 1/2$
gravitino does not feel the Hubble expansion, however, the only 
difference is that now supersymmetry is broken due to the oscillating
scalar field rather than the static vacuum contribution. This is the
reason why instead of the gravitino mass $m_{3/2} \sim 1$ TeV, we now
have $\rho^{1/2}_{\phi}/{\rm M}$.

Substituting Eq.~(\ref{ansatz}), in Eq.~(\ref{lag1}) and after some
algebraic manipulations, for details see \cite{abm}, we derive an
effective Lagrangian

\begin{eqnarray}
\label{app5}
{\cal L}_{\rm eff} \approx e^{G/2} {\partial{G} \over \partial{\phi}}
{\partial{G} \over \partial{\varphi}} \bar{\tilde \phi} \chi_{\rm L}
+{\rm h.c.} \,,
\end{eqnarray}
which is the inflatino coupling in Eq.~(\ref{dom}) to the leading
order. This is the 
most important equivalence
which establishes the fact that, since, for any successful
inflationary model inflaton has to decay, and, so does the inflatino,
the helicity $\pm 1/2$ component of the gravitino must also decay if
the inflaton oscillations is the only viable source of supersymmetry
breaking at that time.  Our result is strictly correct for a single
chiral field responsible for supersymmetry breaking.

\section{Model with a Yukawa coupling to the inflaton}

As a second example we consider a model with a following
superpotential
\begin{eqnarray}
\label{superpot2}
W = {1 \over 2} m {\Phi}^2 + \frac{1}{2}h\Phi \Sigma^2 \,,
\end{eqnarray}
where $\Phi$ contains the inflaton field, which is responsible for the
slow-roll inflation. However, now the inflaton field has an explicit
Yukawa coupling to the matter sector given by the second term in
Eq.~(\ref{superpot2}). Such a superpotential leads to interaction
terms $hm\phi \sigma \sigma$, $h\phi \tilde \sigma \tilde \sigma$,
$h\tilde\phi \tilde \sigma \sigma$, where $\phi $ is the inflaton
field, $\tilde \phi$ is the inflatino, $\sigma $ is a light bosonic
field, and its fermionic partner has been denoted by $\tilde \sigma$.
The estimated rate of the inflaton decaying to fermionic component
$\tilde \sigma~\tilde \sigma$ is given by $\Gamma_{\phi} \sim
(h^2/8\pi)m$.

In general the Yukawa coupling between $\Phi$ and $\Sigma$ multiplets
can also result in the oscillations along the $\sigma$ field. This may
lead to a more complicated
situation where supersymmtery is broken by several multiplets.
However, it is
possible to prevent this provided we require that the $\phi$-induced
mass to the $\sigma$ field is much smaller than the Hubble expansion,
i.e. $h \phi < H$, which implies $h <m/{\rm M}$. We note that this
will also insure that $\sigma$ and 
$\tilde{\sigma}$ are not produced via parametric resonance.  A viable
choice of parameters which can lead to an inflationary
paradigm is; $m =
10^{13}$ GeV, and, a small Yukawa coupling $h = 10^{-7}$, which
ensures that at late stages of the 
inflaton oscillations, ${\phi /{\rm M}} \leq 10^{-14}$, the inflaton
is decaying perturbatively.

We may now repeat the same
analysis as in the previous case which eventually leads to (for
details see \cite{abm})

\begin{eqnarray}
\label{app9}
{\cal L}_{\rm eff} \sim h {\sigma}^{*}\bar{\tilde\phi}\tilde\sigma_{\rm R}
+ {\rm h.c.}\,.
\end{eqnarray}
This reinsures our earlier claim that the equivalence between the
helicity $\pm 1/2$ gravitino and the goldstino is viable at late times
of the inflaton oscillations. This equivalence is not only important
for studying the production of the helicity $\pm 1/2$
components of the gravitino, but also describing the decay of the
helicity $\pm 1/2$ gravitino.

\section{Models with several multiplets}

Once we invoke more than one sectors, and treat them at equal level,
the problem of gravitino production becomes more complicated. This
problem has been addressed in Refs.~\cite{giudice,kallosh1} to
some extent. In this
case it has been realized that the goldstino is a linear combination
of all the fermions, and as a result, even if we use the
goldstino-gravitino equivalence we cannot in general guarantee that a
major contribution to the goldstino mass is coming from the fermionic
component of the inflaton. Interesting question would be to
address a problem where there exists a hidden sector which is
responsible for supersymmetry breaking in that sector, and also
responsible for mediating supersymmetry breaking gravitationally to
the observable sector. In such a case the gravitino will have an
effective mass $\sim {\cal O}({\rm TeV})$ at present vacuum.  
So, keeping this in mind we may consider a simple toy model with a 
following superpotential
\begin{eqnarray}
\label{newsuperpot}
W = {1 \over 2} m_{1} {\Phi}^2 + m_{2}^2[Z+ (2 - \sqrt{3}){\rm M}]\,,
\end{eqnarray}
where $\Phi$ and $Z$ are inflaton and Polonyi multiplets respectively.
We assume that $\phi$ field is responsible for inflation, so we set
$m_{1} = 10^{13}$ GeV to produce adequate density perturbation, while
setting $m_{2} = 10^{11}$ GeV for giving an effective mass to the
gravitino around ${\cal O}({\rm TeV})$. An interesting discussion
regarding this model has been sketched in Ref.~\cite{kallosh1}.

Now one derives a
set of coupled equations for the helicity $\pm 1/2$ gravitino and
other fermionic degrees of freedom \cite{giudice,kallosh1}. It has
been shown in Ref.~\cite{kallosh1}, that in a global supersymmetric
limit, this set of equations is reduced to a coupled set of equations
for the goldstino and the transverse combination of the fermions.
For the above superpotential Eq.~(\ref{newsuperpot}), the inflaton and
the Polonyi sectors have only gravitational interactions. The
fermionic components $\tilde{\phi}$ and $\tilde{z}$ have masses
$m_{1}$ and zero respectively in the global supersymmetric limit.  The
goldstino in this model is a linear combination of the fermionic
components from both the sectors. As long as supersymmtery breaking is
dominated by the inflaton field, the helicity $\pm1/2$ gravitinos 
essentially behave as inflatino. Then the helicity $\pm 1/2$
gravitinos produced during preheating will essentially decay because
they are essentially the inflatino components and so their couplings
are determined in the same fashion as that of the inflaton.

However, the energy density in the inflaton sector is decreasing 
in time, and, when the Hubble expansion $\sim H < {\cal O}(\rm TeV)$, the
$\tilde{z}$ component dominates the goldstino. Usually, the mixing between
the inflatino and $\tilde{z}$ is minimal and Planck mass suppressed, so,
the fermions which are produced during preheating will decay again in 
the form of inflatino and cause no trouble for nucleosynthsis. Once
$z$ field starts oscillating at $H \approx {\cal O}({\rm TeV})$, supersymmetry 
is broken by the oscillations in $z$ direction also, and, as a 
result gravitinos can as well be excited. One may suspect 
that the late production of the helicity $\pm 1/2$ gravitinos will 
dominate and the problem of gravitino decay still persists. The suspicion is
not fully correct because the number density of the helicities $\pm 1/2$
and $\pm 3/2$ are more or less equal now. This is because the only
time-varying scale is due to time-varying mass of the gravitino $\sim
e^{zz^{\ast}/2{\rm M}^2}|W|/{\rm M}^2$. The presence of the Planck
mass suppression prohibits explosive production of the gravitinos at 
late times.
But, now the problem could be much more severe, because these gravitinos with
both the helicities are produced much later, and their effective
masses are also very small roughly of the order of TeV. This leads to
extremely slow decay rate of these gravitinos which may cause a
problem to the Big Bang nucleosynthesis.
Furthermore, the oscillating Polonyi field leads to an even more 
serious problem, i.e. the moduli problem, of which there is no 
satisfactory way out.

Finally, if the fermionic components mix freely, the inflatinos can be  
converted to $\tilde{z}$ (which is the field eventually eaten by the
gravitino).
This presumably occurs around the time when contributions to
supersymmtery breaking from the inflation sector and the Polonyi
sector become comparable. This problem is
analogue to the neutrino flavor conversion and the relevant question
is to ask the conversion probability. 
As mentioned, we beileve that an effcicient conversion will not take
place for the Polonyi model. An efficient conversion nevertheless results
in a large abundance of $\tilde{z}$ fermion, on top of what
is produced due to oscillations of the Polonyi field. We notice that
if the inflatino decays before 
$H \approx {\cal O}({\rm TeV})$, then the abundance of the 
inflatinos prior to conversion will decrease leading to a smaller 
abundance for $\tilde{z}$ (and consequently helicity $\pm 1/2$
gravitinos) even after an efficient conversion.

\section{Conclusion}

Our main result is that in models with one multiplet the coupling of helicity $\pm 1/2$
gravitinos to the supercurrent leads to the same interactions as that of 
the inflatinos when the amplitude of the inflaton oscillations is 
small $|\phi| \ll {\rm M}$.
Then we have argued that the 
production of helicity $\pm 1/2$ states of the gravitino cannot be
considered as a threat for nucleosynthesis.
The helicity 
$\pm 1/2$ states remember their 
goldstino nature and this is the reason why they are produced very 
efficiently compared to the helicity $\pm 3/2$ states. However, the
same goldstino nature also reults in the decay
of the helicity $\pm 1/2$ gravitino alongwith the inflaton. The
requirement that the inflaton must decay to give a successful
nucleosynthesis, leads to an efficient decay of the goldstino, or, the
helicity $\pm 1/2$ gravitinos. This argument holds perfectly well for a single
chiral superfield where the goldstino is inflatino with some additional
phase. This argument can also be applied to models 
where there are more than one sectors of supersymetry breaking,
provided supersymmetry breaking, provided that the
inflationary scale is much higher than the scale of superymmtery
breaking in the hidden sector. Such a situation can arise if there 
exists a Polonyi field in the hidden sector, which we have briefly discussed.
However, we still lack a complete formal tools to explore all possibilities
such as mixing between the fermionic components of the inflaton sector and the
Polonyi sector. This can in principle change the abundance of the helicity 
$\pm 1/2$ component of the gravitinos and a detailed study is certainly 
required in this direction.

The above 
discussion does not apply to the helicity $\pm 3/2$ gravitinos.  
The production of these states during preheating is always Planck mass 
suppressed and so is their couplings to the matter, 
hence they decay quite late and can be dangerous for nucleosynthsis
\cite{anu}.  


\Acknowledgements
The work of R.A. was supported by
``Sonderforchschungsbereich 375 f$\ddot{\rm u}$r
Astro-Teilchenphysik'' der Deutschen Forschungsgemeinschaft.

\end{document}